\documentclass[journal=jacsat,manuscript=article]{achemso}
\usepackage[version=3]{mhchem} 
\usepackage{natbib}
\setkeys{acs}{maxauthors=10, etalmode=truncate}
\usepackage{graphicx}
\usepackage{epsfig}
\usepackage{booktabs}
\usepackage{dcolumn}
\usepackage{bm}
\usepackage[utf8]{inputenc}
\usepackage[T1]{fontenc}
\usepackage{mathptmx}
\usepackage{siunitx}
\usepackage{float}
\usepackage{gensymb}
\usepackage{multirow}

\author{Samuel J. Hall}
\affiliation{Department of Chemistry, University of Warwick, Gibbet Hill Road, Coventry, CV4 7AL, United Kingdom}
\alsoaffiliation{MAS Centre of Doctoral Training, Senate House, University of Warwick, Gibbet Hill Road, Coventry, CV4 7AL, United Kingdom}
\author{Benedikt P. Klein}
\affiliation{Department of Chemistry, University of Warwick, Gibbet Hill Road, Coventry, CV4 7AL, United Kingdom}
\alsoaffiliation{Diamond Light Source, Harwell Science and Innovation Campus, Didcot, OX11 0DE, United Kingdom}
\author{Reinhard J. Maurer}
\email{r.maurer@warwick.ac.uk}
\affiliation{Department of Chemistry, University of Warwick, Gibbet Hill Road, Coventry, CV4 7AL, United Kingdom}

\title[NEXAFS]
  {Characterizing Molecule-Metal Surface Chemistry with Ab-Initio Simulation of X-ray Absorption and Photoemission Spectra}


\begin{document}

\begin{tocentry}
\includegraphics{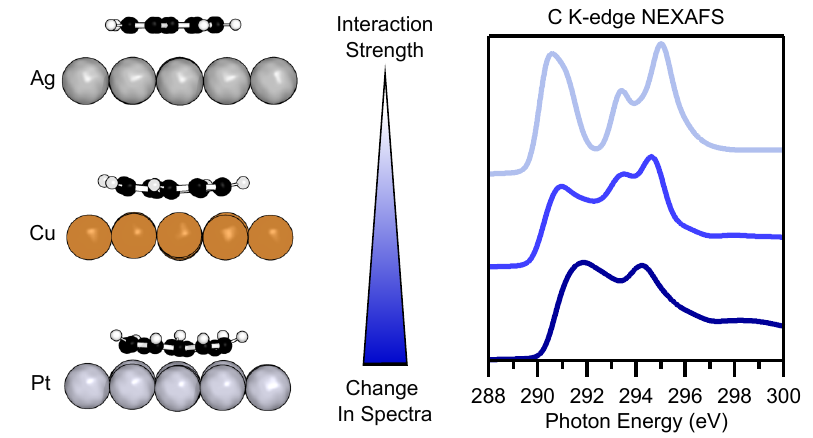}
\end{tocentry}

\begin{abstract}
X-ray photoemission and x-ray absorption spectroscopy are important techniques to characterize chemical bonding at surfaces and are often used to identify the strength and nature of adsorbate-substrate interactions. In this study, we judge the ability of x-ray spectroscopic techniques to identify different regimes of chemical bonding at metal-organic interfaces. To achieve this, we sample different interaction strength regimes in a comprehensive and systematic way by comparing two topological isomers, azulene and naphthalene, adsorbed on three metal substrates with varying reactivity, namely the (111) facets of Ag, Cu, and Pt. Using density functional theory, we simulate core-level binding energies and x-ray absorption spectra of the molecular carbon species. The simulated spectra reveal three distinct characteristics based on the molecule-specific spectral features which we attribute to types of surface chemical bonding with varying strength. We find that weak physisorption only leads to minor changes compared to the gas-phase spectra, weak chemisorption leads to charge transfer and significant spectral changes, while strong chemisorption leads to a loss of the molecule-specific features in the spectra. The classification we provide is aimed at assisting interpretation of experimental x-ray spectra for complex metal-organic interfaces.

\end{abstract}


\section{Introduction}

The strength and nature of interactions at hybrid organic-inorganic interfaces influence the charge transport across the interface. This in turn controls the performance of organic electronic devices, e.g. organic light-emitting diodes \cite{burroughes_light-emitting_1990,forrest_path_2004} or organic field effect transistors \cite{yamaguchi_terazulene_2016,xin_application_2017}. To gain insight into the fundamental mechanisms of the interaction at the interface, model systems consisting of organic molecules adsorbed on single-crystal metal surfaces are often studied using surface science techniques \cite{gottfried_quantitative_2016}. Here, X-ray core-level spectroscopies such as X-ray photoelectron spectroscopy (XPS) and near edge X-ray absorption fine structure (NEXAFS) spectroscopy represent effective tools to characterize the structure and electronic structure of the investigated model systems \cite{stohr_nexafs_1992,bunker_introduction_2010}.

However, the interpretation of the experimental spectra can be highly challenging. Often a large number of unoccupied states contribute to the spectra and overlap significantly. This complicates the assessment of how core levels of different atoms (in XPS and NEXAFS) and different valence states (in NEXAFS) contribute to the measured spectra. Furthermore, without any further atomic-level information on the adsorption structure and electronic properties of the interface, the spectra cannot be connected to important quantities that relate to the nature of the molecule-surface bond, such as the adsorption energy and height, and more conceptual quantities such as the magnitude of charge transfer, and the hybridization between the electronic states originating from surface and molecule, respectively.

First-principles core-level spectroscopy simulations support the interpretation of experimental spectra and are able to disentangle spectra into individual transitions between core and valence states of the system. The methodology in this work is based on core-level constrained density functional theory (DFT) which has been applied before to similar problems and was shown to provide a robust approach for core-level spectroscopy simulations of 1s states in organic molecules \cite{diller_interpretation_2017,klein_nuts_2021}. Application of this method enabled a detailed understanding of the adsorption geometry, chemical bonding and electronic structure \cite{nilsson_probing_2000,morin_chemisorption_2004,fronzoni_density_2012}.

The interaction between a molecule and a metal surface is dependent on the electronic and geometric structure of both participants. On the side of the metal, the reactivity of the substrate can be modified by changing its elemental composition while maintaining the same crystal structure and surface orientation. Noble and coinage metal surfaces with a (111) surface orientation are commonly used as model substrates for fundamental studies. For our work, we chose the three metal substrates Ag(111), Cu(111), and Pt(111). Within these, the reactivity of the metal surface increases from Ag(111) to Cu(111) to Pt(111), as can be directly inferred from the d-band model of surface chemical bonding \cite{hammer_why_1995,pettersson_molecular_2014}. 

On the side of the organic molecule, a wide range of options to tune reactivity exist. The structural variety of organic molecules is almost infinite and minor structural changes can lead to large changes in reactivity. Here, we chose two simple aromatic hydrocarbons, azulene (Az) and naphthalene (Nt). These two molecules are an isomeric pair of bicyclic aromatic hydrocarbons and only differ by the topology of their aromatic system. Naphthalene consists of an alternant 6-6 ring structure and azulene of a nonalternant 5-7 ring structure (see insets in Fig.~\ref{fig:gas_spectra}a) \cite{mallion_golden_1990}. This topological difference between azulene and naphthalene has a large influence on the molecular properties. Solutions of azulene show a brilliant blue color and azulene has a substantial dipole moment while naphthalene is colorless and possesses no dipole moment \cite{lide_crc_2010,klein_topology_2021}. The ability of the two molecules to interact with metal surfaces is also strongly influenced by their different topologies.

A series of recent publications have produced a comprehensive picture of the surface bond of naphthalene and azulene adsorbed onto Cu, Ag and Pt (111) surfaces \cite{klein_molecular_2019,klein_molecule-metal_2019,kachel_chemisorption_2020,klein_enhanced_2020}. The thorough characterization in the literature includes the eludication of the adsorption geometry, energetics and electronic properties by means of various experimental techniques such as near incidence X-ray standing wave (NIXSW), \cite{klein_molecular_2019,klein_molecule-metal_2019} low-energy electron diffraction (LEED), \cite{klein_molecular_2019} XPS, \cite{klein_molecular_2019,klein_molecule-metal_2019,klein_enhanced_2020} NEXAFS, \cite{klein_molecular_2019,klein_molecule-metal_2019,klein_enhanced_2020} temperature programmed desorption (TPD), \cite{klein_molecular_2019,kachel_chemisorption_2020} and single-crystal adsorption calorimetry (SCAC), \cite{klein_enhanced_2020} all combined with DFT simulations. The two molecules and three surfaces represent six interface models that are comprehensively characterized in terms of their structure and electronic properties, which makes them ideally suited for the investigation of how the nature of the respective molecule-metal interaction is reflected in core-level spectroscopic signatures. 

In this manuscript, we build on previously published work on the six molecule-metal interface models \cite{klein_molecular_2019,klein_molecule-metal_2019,klein_enhanced_2020} and present a comprehensive comparative study of first principles simulations of C 1s XPS and C-K edge NEXAFS signatures of azulene and naphthalene adsorbed at Cu(111), Ag(111), and Pt(111). We employ the Delta-Self-Consistent-Field ($\Delta$SCF) \cite{slater_statistical_1972,gunnarsson_exchange_1976,von_barth_dynamical_1980} and Delta-Ionization-Potential-Transition Potential ($\Delta$IP-TP) \cite{stener_density_1995,hu_density_1996,triguero_calculations_1998} methods to characterize and analyze the XP and NEXAFS spectra of these large periodic systems. Using these simulations, we identify three different molecule-metal surface bonding regimes: physisorption, weak chemisorption (one-way charge transfer), and strong chemisorption (two-way charge transfer leading to molecule-metal hybridization), with each regime showing characteristic signatures and changes in the respective spectra compared to the gas-phase data. We expect our findings to be useful to interpret experimental spectral changes for complex hybrid organic-inorganic thin films.

\section{Computational Details}

The structural models used for the spectroscopic simulations in this study were taken from the literature \cite{klein_molecular_2019,klein_molecule-metal_2019,klein_enhanced_2020}. In these publications, the structural optimization was performed using a combination of the PBE functional and the DFT-D3 van der Waals dispersion correction. The reported structures were previously found to be in good agreement with experimental data, which has been summarized in the SI in Figs. S1 and S2. Metal surfaces were modelled as 4-layer slabs with a ($2\sqrt{3}\times2\sqrt{3}$)-R\ang{30} unit cell containing 48 metal atoms in total. More details on the computational settings employed can be found in the relevant literature references \cite{klein_molecular_2019,klein_molecule-metal_2019,kachel_chemisorption_2020,klein_enhanced_2020,hall_computational_2022}.

All core-level calculations in this study are based on previously optimised structures and were performed with the electronic structure software package CASTEP 18.11 \cite{clark_first_2005} which utilizes periodic boundary conditions (PBC). Default on-the-fly generated ultrasoft pseudopotentials and the PBE exchange correlation functional \cite{perdew_generalized_1996} were used throughout. We employ a planewave (PW) cutoff energy of \SI{450}{\electronvolt} and a k-grid of $6\times6\times1$ for all metal surfaces ($1\times1\times1$ for the gas-phase calculations). These values provide a converged potential for carbon and for all three metal surfaces investigated in this work. An electronic convergence criterion for the total density of at least \SI{1d-6}{\electronvolt/atom} was employed. The influence of these parameters has been tested thoroughly in a previous publication and shown to give well converged results \cite{klein_nuts_2021}.

XPS simulations employed the $\Delta$SCF method, \cite{slater_statistical_1972,gunnarsson_exchange_1976,von_barth_local-density_1979} calculating the core-electron binding energy (BE) from the difference in total energy of two singlepoint calculations, one being the ground-state configuration and the second a core-hole excited configuration, where one electron is removed from the 1s orbital. This method is implemented in CASTEP by modifying the pseudopotential definition of the excited atom to include a full core-hole \cite{gao_core-level_2009,mizoguchi_first-principles_2009,klein_nuts_2021}. Such an excited state calculation is carried out for every individual carbon atom in the molecule in order to produce the full XP spectrum.

NEXAFS simulations were performed using the $\Delta$IP-TP method \cite{triguero_calculations_1998,klein_nuts_2021}. The TP approach allows for all transition energies from the 1s state of one atom into all possible unoccupied states to be calculated in a single calculation. Modified pseudopotentials are used again but here include only half a core-hole instead of a full core-hole. The ELNES module \cite{pickard_ab_1997,gao_theory_2008,mizoguchi_first-principles_2009} in CASTEP was used to simulate NEXAFS energies and transition dipole moments. This module performs a total energy SCF calculation followed by a band structure calculation in order to converge the unoccupied states. Inclusion of 800 unoccupied bands was enough to cover the spectral range for all systems. The $\Delta$IP-TP extension of the TP method involves the shift of all transition energies (1s $\rightarrow$ unoccupied states) belonging to each atom according to the XPS binding energies of its 1s electrons, which were obtained by the $\Delta$SCF calculation in the previous step. This ionization potential correction aligns all individual core-level spectra to the same energy frame.

Post-processing of the data was carried out through the use of a dedicated tool, MolPDOS, \cite{maurer_excited-state_2013} which is part of the CASTEP source code. The molecular orbital (MO) projection scheme contained in this tool allows us to estimate what part of the spectrum can be attributed to the electronic states of a reference system. Application of this method for gas-phase molecules allows for a full decomposition of the NEXAFS spectrum in terms of the individual MO states of the molecule. For molecules adsorbed on surfaces, we choose the free standing molecular overlayer (removal of the metal slab in structure) as reference and perform a MO projection to determine the overlap of the electronic states of this reference with the final-states of each transition in the combined system (molecule and metal surface). Our previous work shows this approach to work consistently well to identify molecular contributions to spectra of adsorbed molecules \cite{klein_topology_2021,klein_nuts_2021,diller_interpretation_2017}.

Finally, a pseudo-Voigt broadening scheme was used to simulate experimental broadening effects and therefore convert the calculated transition energies and intensities to simulated spectra resembling experimental data \cite{schmid_new_2014,schmid_new_2015,klein_nuts_2021}. A comprehensive description of how these calculations are conducted has been published previously \cite{klein_nuts_2021}.

The raw input and output files for all calculations have been deposited in the NOMAD repository and are freely available online via https://dx.doi.org/10.17172/NOMAD/2021.06.14-1.

\section{Results}

\subsection{\label{subsec:topological_diff}Effect of molecular topology on XPS and NEXAFS signatures}

\begin{figure}[ht!]
    \includegraphics{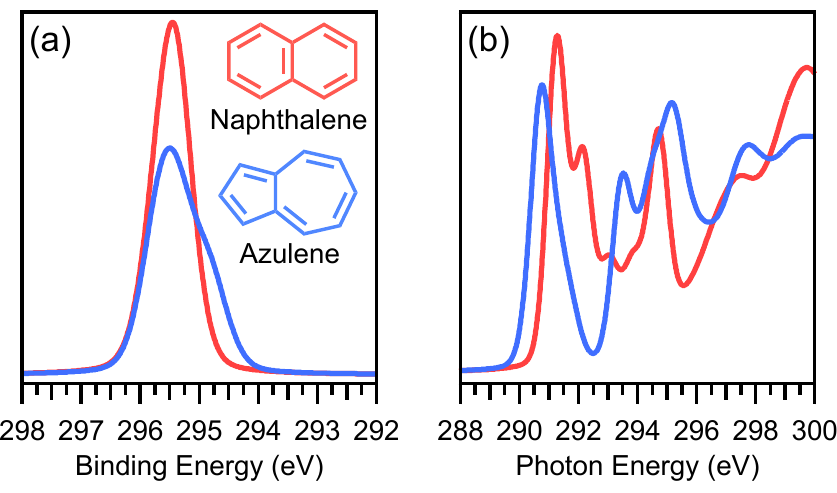}
    \caption{\label{fig:gas_spectra} Comparison of the simulated core electron spectra for the gas-phase structures of molecules naphthalene (red) and azulene (blue). (a) C 1s XP spectra previously reported in Ref.~\citenum{klein_molecular_2019}. (b) C K-edge NEXAFS spectra, showing the total, angle independent spectra \cite{klein_nuts_2021}.}
\end{figure}

First, we will concentrate on how the difference in topology between azulene and naphthalene influences their gas-phase XP and NEXAFS spectra (see Fig.~\ref{fig:gas_spectra}). When a direct comparison of the calculated spectra to experimental data (obtained by gas-phase or molecular crystal measurements) is desired, a global rigid energy shift has to be applied to match the simulation to the experimental energy scale. This correction is necessary due to the frozen core approximation in our pseudopotential plane-wave calculations and the approximations in the employed exchange-correlation functional \cite{klein_nuts_2021}. In this work, however, we will not be comparing directly to experimental data. Therefore all spectra in this publication are shown with the original energy scale as obtained by the calculations.

The topological difference between naphthalene (alternant 6-6 structure) and azulene (nonalternant 5-7 structure) has a strong influence on both the XPS and NEXAFS data. In the XP spectra (Fig.~\ref{fig:gas_spectra}a) a much broader peak is observed for azulene, with a pronounced shoulder visible at lower binding energies. In the NEXAFS spectra (Fig.~\ref{fig:gas_spectra}b) azulene shows the leading edge at lower photon energy and the first peak has a shoulder at higher energy, while naphthalene has its leading edge at higher photon energy and two maxima within the first spectral feature.

\begin{figure}[ht!]
    \includegraphics{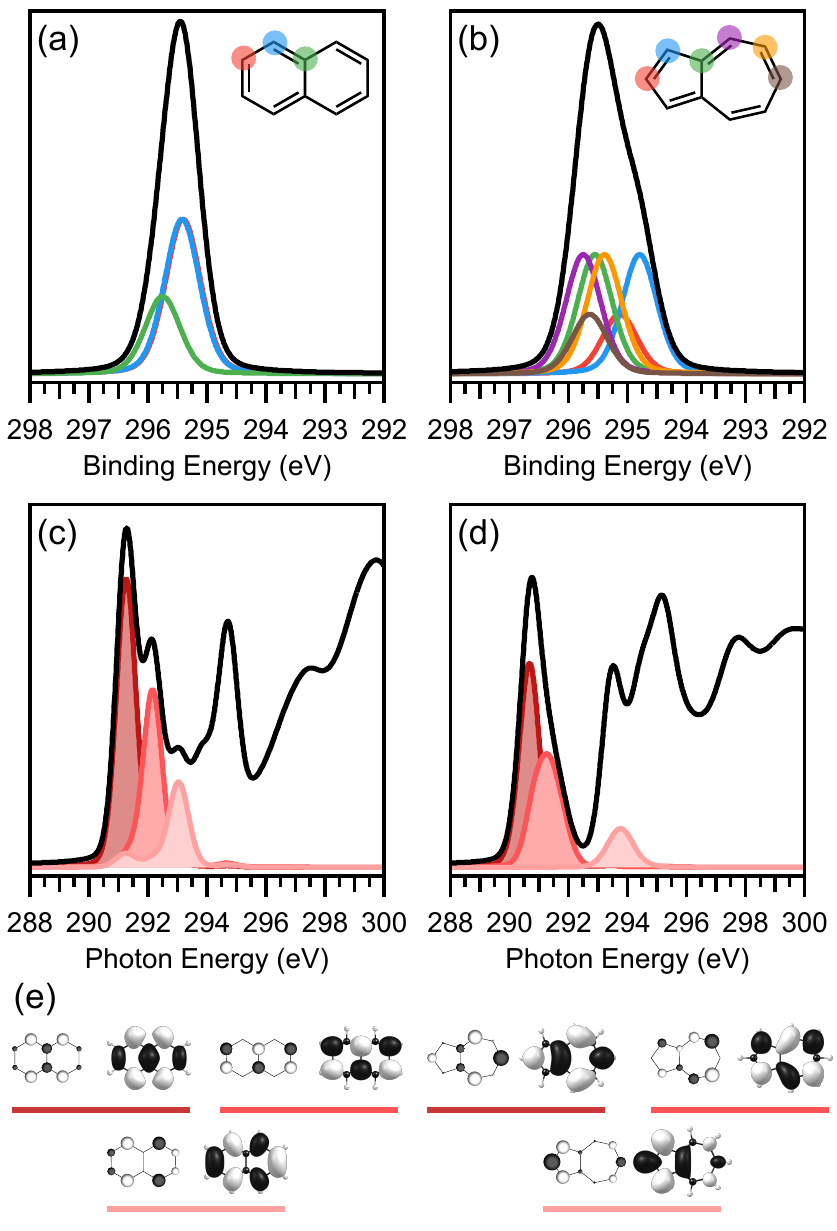}
    \caption{\label{fig:gas_decomp}Initial state decomposition of XPS for gas-phase naphthalene (a) and azulene (b), previously published in Ref.~\citenum{klein_molecular_2019}, showing the C 1s atomic orbital contributions for each colored atom in the corresponding structure. NEXAFS final state decomposition for the total, angle independent spectra \cite{klein_nuts_2021} for naphthalene (c) and azulene (d), showing the spectral contribution of respective molecular orbitals LUMO to LUMO+2 in increasingly lighter shades of red. The corresponding MOs are visualized in (e).}
\end{figure}

The XP and NEXAFS spectra can be analyzed by disentangling the initial-state and final-state contributions to the overall spectrum, as shown in Fig.~\ref{fig:gas_decomp}. The individual contributions of each carbon atom to the total XP spectrum are shown for naphthalene (Fig.~\ref{fig:gas_decomp}a) and azulene (Fig.~\ref{fig:gas_decomp}b). The shoulder of the peak in the azulene spectrum can be attributed to the carbon atoms in the 5-membered ring possessing a lower C 1s binding energy than those in the 7-membered ring, which arises from the strong dipole moment of the molecule and the related inhomogeneous charge distribution. For naphthalene, the C 1s binding energies are similar for all carbon atoms, with only slightly larger values for the bridging carbon atoms, yielding an almost symmetric peak shape.

In contrast to the shown initial-state decomposition of the XP spectra, the decomposition of the NEXAFS spectra (Fig.~\ref{fig:gas_decomp}c,d) is performed according to the final-state contributions to the transitions, i.e. the contribution to the spectrum that arises from all transitions from any 1s state into the lowest unoccupied molecular orbital (LUMO), the LUMO+1 etc. Each colored peak represents all transitions into the same molecular orbital as final-state, no matter what the initial-state, i.e. C 1s orbital, is involved. For naphthalene (Fig.~\ref{fig:gas_decomp}c) we clearly see that the first three peaks in the spectra originate from transitions into the three lowest unoccupied molecule orbitals (LUMO, LUMO+1, LUMO+2). For azulene (Fig.~\ref{fig:gas_decomp}d) the first peak consists of transitions into the first two unoccupied orbitals, with the transition into the LUMO forming the leading edge and the transition into the LUMO+1 forming the shoulder at higher photon energy.

Our simulations show that the topology of the molecular backbone has a pronounced influence on the shapes of the spectral features both in XPS and NEXAFS. The overall shape of the simulated spectra is also in good agreement with experimental data for multilayers of the molecules \cite{klein_molecular_2019,klein_enhanced_2020}. Additionally, the topology of the molecules affects the adsorption of the molecules on metal surfaces, which we discuss in the next section.

\subsection{\label{subsec:metal-organic-intf}Properties of the molecule-metal interfaces}

To sample a wide range of different molecule-metal interactions, the two molecules naphthalene and azulene were combined with three different metal (111) surfaces of increasing reactivity, namely Ag(111), Cu(111), and Pt(111). Comprehensive experimental investigations into these systems are available in the literature and provide a thorough characterization of the molecule-metal interaction based on a wide variety of experimental and theoretical techniques \cite{klein_molecular_2019,klein_molecule-metal_2019,kachel_chemisorption_2020,klein_enhanced_2020}. 

\begin{figure*}[ht!]
    \includegraphics{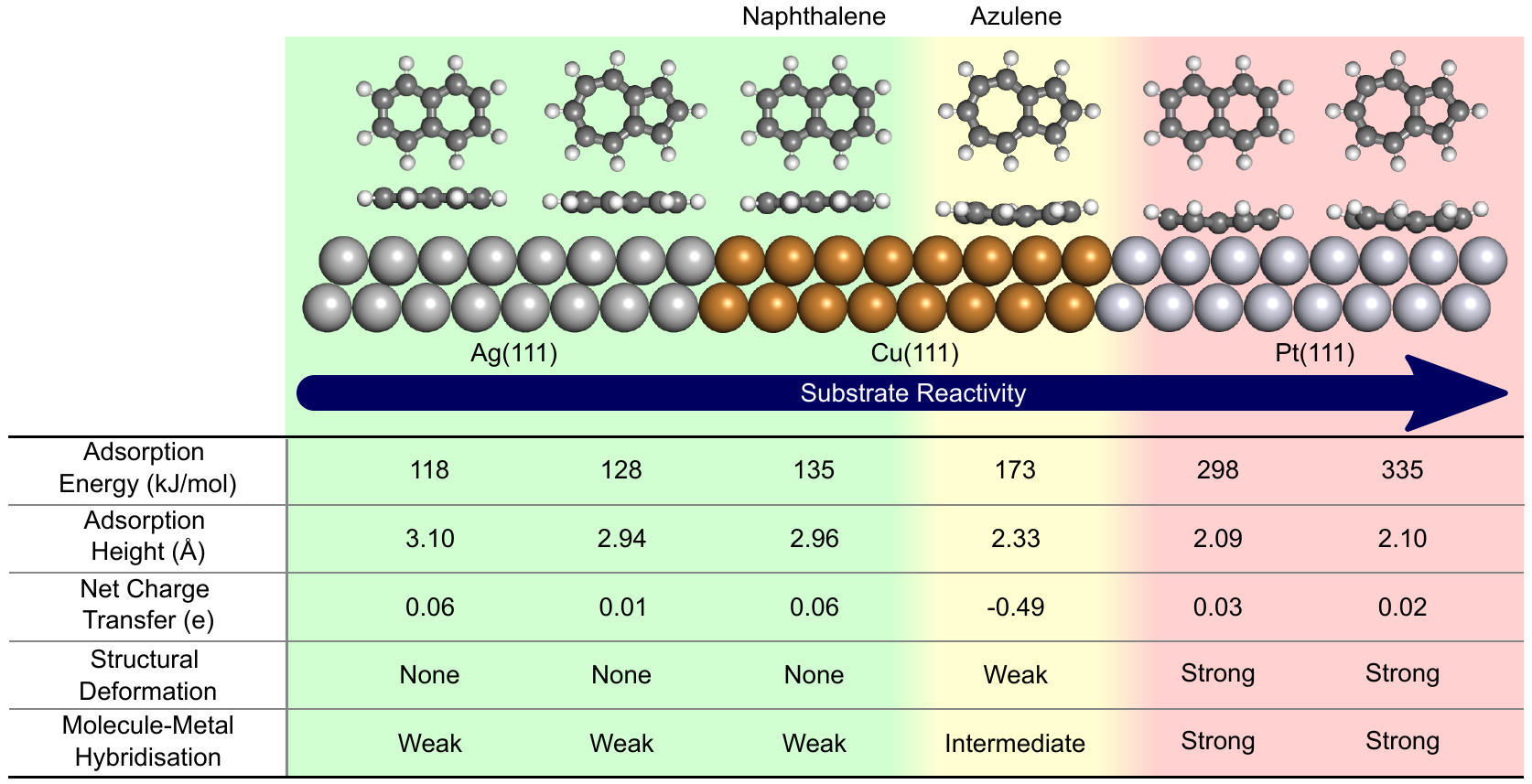}
    \caption{\label{fig:systems}Schematic depiction of the studied six metal-organic interfaces arranged according to increasing interaction strength from left to right. Background color represents the classification of the interaction regime: green, physisorption regime; yellow, weak chemisorption regime; red, strong chemisorption regime. The table shows adsorption energies, adsorption heights, charge transfer, as well as two qualitative criteria for the interaction strength. The presented data is compiled from Refs.~\citenum{klein_molecule-metal_2019,kachel_chemisorption_2020,klein_enhanced_2020} and was obtained with PBE-D3.  The values for the charge transfer are results of the Bader atoms-in-molecules method. Further results of additional charge transfer analysis methods can be found in the same references and are summarized in Table S1 of the SI.}
\end{figure*}

Fig.~\ref{fig:systems} summarizes this information regarding all the metal-organic systems involved, arranged in order of increasing interaction between the molecule and the metal surface as given by the adsorption energy. The top row of Fig.~\ref{fig:systems} shows the structures of naphthalene and azulene, while the second row contains the side-on view of the adsorbed structures to highlight the change in adsorption height and molecule deformation. Below the structures is a table containing a range of additional parameters that can be used to describe the strength of interaction between the molecule and metal surface. As we move across from left to right, the reactivity of the metal increases from Ag to Cu to Pt. Accordingly, the adsorption energy increases and the adsorption height decreases, showing the strengthening of the bond between the molecule and surface. Furthermore, when comparing the two molecules adsorbed on the same metal surface, a stronger interaction is observed for azulene than for naphthalene. The 5-7 nonalternant topology of azulene therefore leads to an increased reactivity at the metal-organic interface \cite{klein_molecular_2019}.

In contrast to the clear trends in adsorption energy and height, the charge transfer between the molecule and surface (as predicted by the Bader charges) paints a more complicated picture. For the first three systems, both molecules on Ag(111) and Nt/Cu(111), little to no charge transfer between the molecule and the metal can be observed. For Az/Cu(111) a substantial amount of charge transfer, \SI{-0.49}{e} from the surface to the molecule, shows a strong interaction between the electronic states of the molecule and metal. But for both molecules adsorbed on Pt(111), the  net charge transfer is vanishingly small even though the values of the adsorption height and energy show a very strong bond and the molecule is strongly deformed upon adsorption. This behavior arises due to a strong electronic interaction based on donation and back-donation of electrons between the metal and the surface. In the case of Az/Cu(111), with an adsorption height of \SI{2.33}{\angstrom}, hybridization is not too strong and charge transfer occurs mainly due to Fermi level pinning. This results in exclusive electron transfer from the Cu(111) surface to the diffuse unoccupied states of the molecule. Because both molecules are much closer to the Pt(111) surface, the more compact occupied electronic states of the molecule are now also available for electron back-donation into the unoccupied states of the metal surface. The result is a two-way charge transfer with donation and back-donation, comparable to the situation in many organic transition metal complexes as described by the Dewar-Chat-Duncanson (DCD) model. \cite{mingos_historical_2001,chatt_586_1953,klein_enhanced_2020} This results in an almost net zero overall charge as shown by the charge partitioning analysis.

Finally, the six systems can be compared by two qualitative parameters: (1) structural deformation and (2) hybridization of the electronic states of molecule and metal. The deformation describes how much the molecule and the surface have changed in the adsorbed state compared to their relaxed structures. Again, the first three systems show only weak disturbance, with a planar molecule and all molecular C-C bond lengths unchanged \cite{klein_molecular_2019,klein_molecule-metal_2019}. For Az/Cu(111), we classify the deformation as weak, because a slight buckling is now present in the molecule and bond lengths have started to change noticeably \cite{klein_molecular_2019,klein_molecule-metal_2019}. However, when adsorbed on Pt, the structural deformation of both molecules becomes significant. Here, the molecules are strongly buckled, with the hydrogen atoms pointing upwards away from the surface. For all carbons, the bond angles and bond lengths are more in agreement with aliphatic sp$^3$ geometry than the aromatic sp$^2$ geometry of the free molecules \cite{klein_enhanced_2020}. The pattern observed for the deformation is also visible in the hybridization between electronic states of the molecule and metal surface, which can be assessed by analysis of the projected density-of-states (DOS) of the system. Only weak hybridization is present for the first three systems with the MO signatures only weakly broadened \cite{klein_molecular_2019,klein_molecule-metal_2019}. For Az/Cu(111), projected MO signatures in the DOS are further broadened and exhibit some level of splitting \cite{klein_molecular_2019,klein_molecule-metal_2019}. For both molecules on Pt(111), molecular resonances are strongly hybridized with the metal covering a wide range of energies across the DOS \cite{klein_enhanced_2020}. The topic of hybridization will be discussed in detail in the context of the core-level spectra.

All the information laid out above is supported by the experimental data in the literature and leads us to identify three different regimes of interaction strength between the molecules and the metal surface. Type I (shaded in green in Fig.~\ref{fig:systems}) includes Nt/Ag(111), Nt/Cu(111) and Az/Ag(111) and represents the weakest level of interaction, which we refer to  as physisorption. Type II (yellow) contains solely Az/Cu(111), and is best described as weak chemisorption. The main indicator of this regime is the presence of a one-way charge transfer between metal and molecule that arises from Fermi level pinning of the azulene LUMO. Finally, Type III (red) encompasses Nt and Az on Pt(111), which we designate as strongly chemisorbed systems that exhibit strong molecular deformation, large adsorption energies and adsorption heights consistent with atomic covalent radii. In the following sections we will discuss the characteristic spectral features in XP and NEXAFS spectroscopy associated with the three regimes of molecule-metal interaction strength.


\subsection{\label{subsec:classI}Physisorption Regime}

\begin{figure*}[ht!]
    \includegraphics{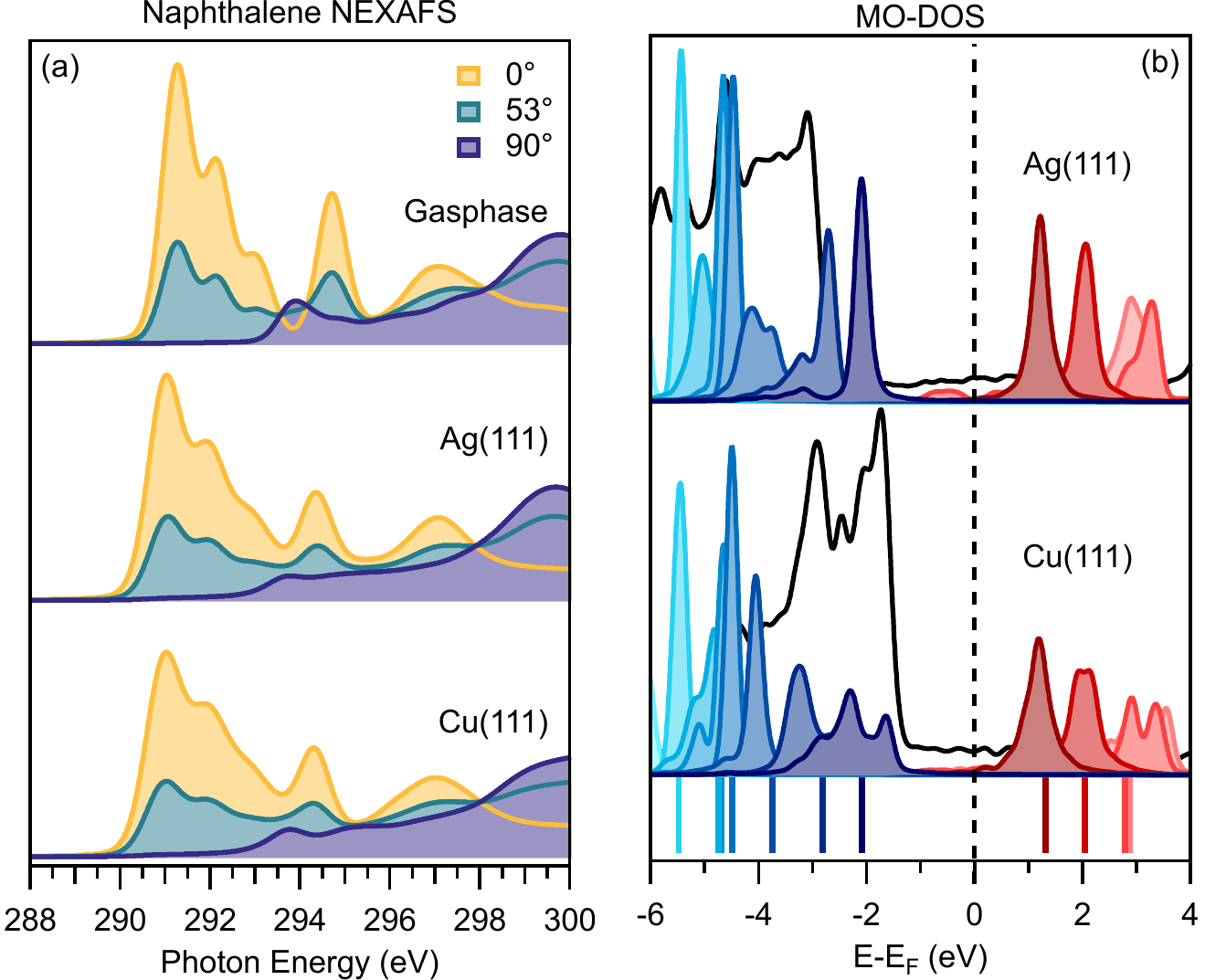}
    \caption{\label{fig:np_class1}Comparison of the (a) NEXAFS spectra and (b) DFT molecular orbital projected density of states. The NEXAFS spectra presented are for naphthalene in the gas-phase (top), adsorbed on Ag(111) (middle), and adsorbed on Cu(111) (bottom). For each system, three different incidence angles are shown (\ang{0}, yellow; \ang{53}, turquoise; \ang{90}, blue). MO-DOS shows the total DOS (black line) against scaled MO projections of naphthalene adsorbed on Ag(111) (top), and on Cu(111) (bottom). The Fermi level is shown as a vertical dashed line. Contributions shaded in blue represent projection onto occupied states, while contributions shaded in red represent unoccupied states. Lighter shades of these colors show states lower or higher in energy, respectively. Colored lines at the bottom of the graph represent gas-phase orbital energies shifted by \SI{3.31}{\electronvolt} to account for the work function of the adsorbed system (see text for details). All data was previously published in Ref.~\citenum{klein_molecule-metal_2019}, except for the \ang{0} NEXAFS spectra.}
\end{figure*}

The physisorption regime describes the weakest level of interaction between the molecule and metal surface and includes Nt/Ag(111) and Nt/Cu(111) in Fig.~\ref{fig:np_class1} and Az/Ag(111) in Fig.~\ref{fig:class2}. The XPS spectrum of naphthalene in the gas-phase and adsorbed onto Ag(111) and Cu(111) are virtually identical as can be seen in Fig. S3 of the SI. In the case of azulene the shoulder present in the gas-phase XP spectrum (see Fig~\ref{fig:gas_spectra}) disappears when adsorbed on Ag(111), which presents a significant change.  However, the disappearance of the shoulder has been found to be the cause of an intrinsic error in common Density Functional Approximations which causes spurious charge transfer between the molecule and metal \cite{hall_self-interaction_2021}. We have previously discussed this issue and how the spurious charge transfer can be addressed with a correction \cite{hall_self-interaction_2021}. In Fig.~\ref{fig:class2}a, we present the results of the unchanged $\Delta$SCF method together with the corrected XP spectra, labelled Ag(111)+U. The correction yields spectra where the shoulder is recovered which is in better agreement with experiment and gas-phase data (see Fig~\ref{fig:gas_spectra}). 

For naphthalene in the gas-phase and adsorbed on Ag(111) and Cu(111),  the correction was not necessary and the resulting simulated NEXAFS spectra are presented in Fig.~\ref{fig:np_class1}a. Similar to the (in the case of azulene: corrected) XPS, it is apparent that the adsorption of naphthalene on Ag(111) and Cu(111) does not introduce major differences in the NEXAFS spectra. The only difference observed is a weak additional broadening across the spectrum, whereas peak positions and intensities are not strongly affected. The inspection of the spectra with different polarization angles also shows that the dichroism from the gas-phase spectra is retained, i.e. the dependence of the signal with respect to the incident light polarization. In all cases, the signals from the first two peaks diminish almost completely to zero for the normal incidence case (\ang{90}). The fact that these peaks vanish at normal incidence is due to the orientation of the unoccupied states with respect to the incident radiation. As the LUMO, LUMO+1, and LUMO+2 are $\pi^*$ orbitals, for which the molecular plane contains a nodal plane, they can only interact with light that has a polarization contribution perpendicular to the molecular plane.

To show the reason for the lack of change in the NEXAFS spectra, the molecular orbital projected density-of-states (MO-DOS) is shown in Fig.~\ref{fig:np_class1}b. For both systems, there is a clear gap between occupied molecular states (in shades of blue) and unoccupied states (in shades of red). All occupied states remain below the Fermi level (dashed line) and all unoccupied states stay above it, which indicates negligible charge transfer between molecule and surface. The colored lines at the bottom of  Fig~\ref{fig:np_class1}b are the molecular states of the gas-phase molecule. The work function of the metal surface was accounted for by aligning the molecular orbitals of lowest energy for gas-phase molecule with the ones for the adsorbed molecules. The difference in alignment for Ag(111) and Cu(111) was only \SI{0.02}{\electronvolt}, therefore the visualization is valid for both systems. 

While there is little shift in the positioning of the orbitals on either surface, some increased broadening can be seen when the molecule is adsorbed on Cu(111). This finding can be attributed to the higher reactivity of the metal which has been previously identified through the slightly higher adsorption energy and smaller adsorption height (see Fig~\ref{fig:systems}). Overall, however, minimal influence of the adsorption on the NEXAFS spectra and DOS are in accordance with the previously discussed lack of charge transfer and electronic hybridization. Therefore, in the physisorbed bonding regime, measurements of gas-phase and multilayer spectra should closely reflect spectra in the metal-adsorbed monolayer

\subsection{\label{subsec:classII}Weak Chemisorption Regime}

\begin{figure*}[ht!]
    \includegraphics{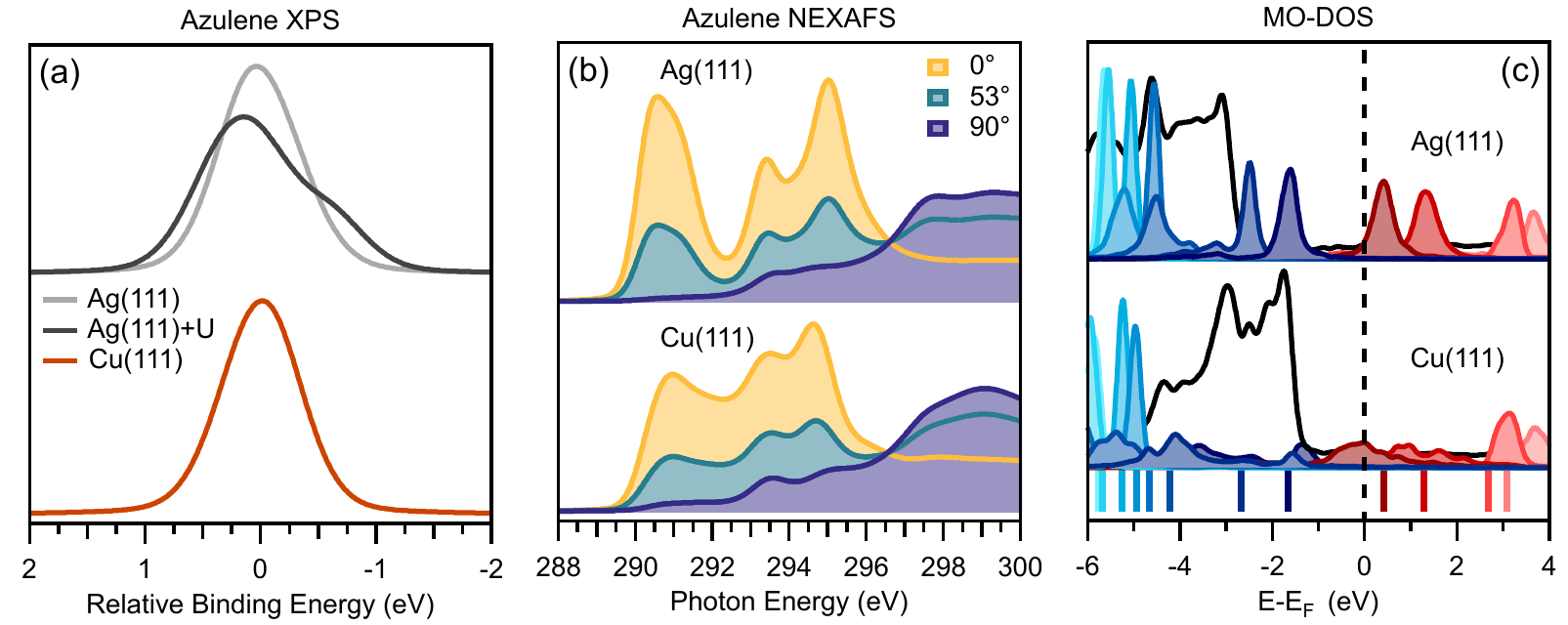}
    \caption{\label{fig:class2}Comparison between physisorbed Az/Ag(111) and weakly chemisorbed Az/Cu(111) (a) XPS simulations, with an additional +U(MO) corrected spectrum for Az/Ag(111), aligned with the average binding energy set to zero, to account for spurious charge transfer \cite{hall_self-interaction_2021}. (b) NEXAFS simulations at three different incidence angles. (c) DFT density of states (DOS), with scaled MO projections. The total DOS is shown in black and the Fermi level as a dashed vertical line. Contributions in blue represent projected occupied states, while contributions in red represent unoccupied states. Lighter shades of these colors show states lower or higher in energy, respectively. The colored lines at the bottom of the graph represent the gas-phase orbitals energies shifted by \SI{3.23}{\electronvolt} to account for the work function of the adsorbed system (see text for details). NEXAFS spectra (except for the \ang{0} NEXAFS spectra) have previously been published in Ref.~\citenum{klein_molecule-metal_2019}. The Ag(111)+U XPS spectra have previously been reported in Ref.~\citenum{hall_self-interaction_2021}.}
\end{figure*}

The weak chemisorption regime in our list of six systems only consists of Az/Cu(111). The change in bond lengths and loss of planarity of the molecule due to adsorption impact both the XP and the NEXAFS spectra, as is apparent in the direct comparison between physisorbed Az/Ag(111) and weakly chemisorbed Az/Cu(111), (Fig.~\ref{fig:class2}).

When discussing the XP spectra (Fig.~\ref{fig:class2}a), a special point has to be mentioned. For Az/Ag(111), the intrinsic DFT error that leads to a spurious charge transfer has to be taken into account \cite{hall_self-interaction_2021}. Otherwise, the simulated spectrum based on the conventional $\Delta$SCF approach (light gray in Fig.~\ref{fig:class2}a) shows a symmetric peak which has lost the shoulder present in the gas-phase spectrum (Fig.~\ref{fig:gas_spectra}). This change in peak shape disagrees with the experimental data for Az/Ag(111), where the shoulder is present \cite{klein_molecule-metal_2019}. Comparison of the corrected XPS spectrum for Az/Ag(111) and the one for Az/Cu(111) shows that the increased interaction from physisorption to weak chemisorption has a noticeable effect. The interaction with the Cu(111) surface results in charge transfer into the LUMO of the molecule, in turn leading to reduced relative shifts between the individual carbon atoms. As a consequence, the contributions from the atoms in the 5- and 7-membered rings are no longer distinguishable, and a single, symmetric peak is present. This is the case both in experiment and in our simulations, which means that, contrary to Az/Ag(111), no qualitative discrepancies arise for Az/Cu(111) between experiment and the conventional $\Delta$SCF simulations.

The polarization dependent NEXAFS spectra (Fig.~\ref{fig:class2}b) also show a significant difference between azulene adsorbed on the Ag(111) or the Cu(111) surface. The spectra were calculated using the XPS binding energies obtained with the conventional $\Delta$SCF method. The clear change when going from Ag(111) to Cu(111) is seen by an overall broadening and a significant diminishing of intensity as well as a shift to higher photon energy for the first peak, whilst the second and third are slightly shifted to lower energies (see Table S2 of the SI for the relative energies of the first three peaks). When compared to the peak intensities of the gas-phase spectra, and normalized with respect to peak 3, the first peak lowers to \SI{68}{\percent} on silver and to only \SI{53}{\percent} intensity on copper. On the other hand, the intensity of the 2nd peak remains about the same with \SI{97}{\percent} when adsorbed on silver and has even an increased relative intensity to \SI{127}{\percent} when adsorbed on copper (see the absolute intensities for each peak in Table S2 and the normalized values in Tables~S3 and S4 of the SI). In literature, the diminishing of the first peak is often attributed to charge transfer between the electronic states of the molecule to the metal but this is virtually impossible to prove from experiment alone \cite{puglia_physisorbed_1995,medjanik_orbital-resolved_2012}.

Comparing the NEXAFS spectra simulated with different incidence angles in Fig~\ref{fig:class2}b, we also see a reduction in dichroism on the copper surface, with a larger amount of residual intensity remaining for the first peak at \ang{90} incidence angle. This intensity should be vanishing, because the symmetry selection rules forbid an excitation into the LUMO (and other $\pi$* states) when the dipole of the MO is perpendicular to the polarization vector of the incident light at \ang{90}. The presence of residual intensity proves that the selection rules are not valid anymore, because the molecule is slightly deformed and the electronic states of the molecule are already hybridized with each other and with metallic states.

The MO-DOS for azulene on the two metal surfaces shows a clear difference between the two interaction regimes (Fig.~\ref{fig:class2}c). When azulene is adsorbed on Ag(111), the band gap between the HOMO and LUMO is preserved with the HOMO being below the Fermi level (black dashed line) and LUMO still above. However, when azulene is adsorbed on Cu(111), both the HOMO and LUMO exhibit a different shape in the DOS, they are broadened and smeared out over a wide energy range. Furthermore the LUMO is now partly below the Fermi level and therefore partially occupied. These findings are a clear indication of hybridization and charge transfer caused by substantial interaction between the molecule and the copper surface, which was not the case for the silver surface. 

The MO-decomposed NEXAFS spectra (see Fig. S5 of the SI) shows the relative contribution of the ground-state MOs projected onto the NEXAFS spectrum. For the gas-phase projected spectra shown in Fig.~\ref{fig:gas_decomp} the sum of the orbitals equals the total spectra. This is not true anymore for the molecule adsorbed on a metal surface, here the contributions do not sum up to the total spectrum. The intensity breakdown of the orbitals with respect to the overall NEXAFS spectra therefore cannot be described quantitatively. However, the differences in the MO-contributions when comparing different systems still offer some qualitative insight. For azulene adsorbed on Cu(111) (see Fig. S5 of the SI), it is apparent that the transitions into the first two unoccupied orbitals (LUMO and LUMO+1) diminish greatly compared to when adsorbed on Ag(111). Furthermore, those two transitions now possess a lower intensity than the transition into the higher LUMO+2 state. This correlates with the hybridization seen in the MO-DOS. For Az/Cu(111), the reduction of the leading edge peak can therefore be directly associated with the adsorption induced change in the LUMO and LUMO+1 and metal-to-molecule charge transfer.


\subsection{\label{subsec:classIII}Strong Chemisorption Regime}

The third and final interaction type is the strong chemisorption observed for both azulene and naphthalene adsorbed on the Pt(111) surface. The XP spectrum for Az/Pt(111) (see Fig.~S3 of the SI) shows the lower energy shoulder disappearing from the spectrum, as was already observed when adsorbed on Cu(111). Again, this is due to the interaction between the molecule and metal eliminating the BE differences of the individual carbon atoms. Looking at the naphthalene spectrum, not much change is seen between the spectrum when adsorbed on Pt(111) and the physisorbed state observed when adsorbed on Ag(111) and Cu(111), meaning that all carbons are equally influenced by the interaction with the substrate. 

\begin{figure*}[ht!]
    \includegraphics{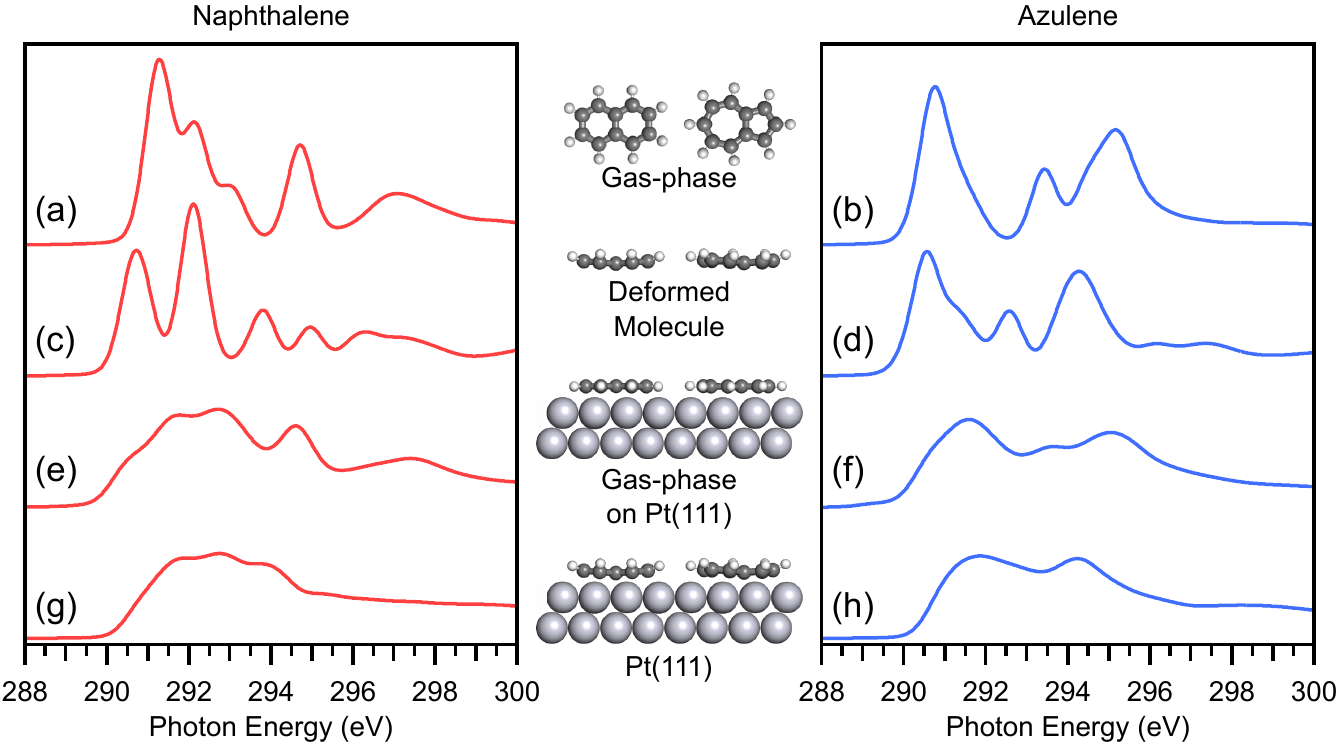}
    \caption{\label{fig:class3}Simulated NEXAFS spectra, at \ang{0} incidence angle, for naphthalene (left) and azulene (right) in different configurations. (a,b) is the free gas-phase molecule, (c,d), the deformed molecule without the metal surface, (e,f), the planar gas-phase molecule at the Pt(111) site  and (g,h) the molecule adsorbed on the Pt(111) surface.}
\end{figure*}

The increased interaction strength has a strong effect on the NEXAFS simulations for both molecules adsorbed on Pt(111) (see Fig.~\ref{fig:class3}g,h). The spectra have virtually lost all connection to the unoccupied molecular states previously seen and any residual features are highly broadened. Both spectra are now similar in shape and indeed almost indistinguishable from one another  because all identifiable features which arose from the difference in molecular topology (see Fig.\ref{fig:gas_spectra}) are diminished. This is a consequence of the strong hybridization between the electronic states of molecule and metal. The MO-projected NEXAFS shows how the contribution of the molecular states to the overall spectra have been reduced even further compared to the weak chemisorption regime and that a small contribution stemming from excitations into the former HOMO is now present above the Fermi level due to its partial deoccupation (Fig.~S6 of the SI). In the MO-DOS of both molecules on Pt(111) (Fig.~S7 of the SI) strong broadening of the MOs over a wide energy range as a result of hybridization is apparent. Here we can clearly see the donation/back-donation that results in the negligible net charge transfer as the HOMO and LUMO have spread out to both sides of the Fermi level leading to them being partly depopulated and populated, respectively. As a consequence, the electronic states of the molecules are deeply integrated into the metallic bands of the Pt surface.

Studying the polarization dependence of the NEXAFS spectra, we observe a massive reduction in the dichroism of the NEXAFS signal (Fig.~S8g,h), suggesting an almost complete breakdown of the symmetry selection rules for the adsorbed molecule. The molecular electronic structure is now deeply embedded in the metallic band structure and the molecular orbitals of the gas-phase molecule cease to be meaningful descriptors for the excitation process.

Due to the strong bond to the surface, the molecule experiences both (1) a significant adsorption induced deformation and (2) a strong hybridization between its molecular states and the electronic states of the surface. Therefore it is an interesting task to investigate which of these effects is dominant for the adsorption-induced change observed in the spectra. The nature of this problem makes it inaccessible to experimental analysis and only solvable by an in-depth theoretical study. In the following we will use calculations for several hypothetical model structures to provide insight into the causes of the spectral change due to adsorption. 

Fig.~\ref{fig:class3} shows NEXAFS calculations for several structures related to Az and Nt adsorbed on Pt(111). First are the gas-phase spectra showing the already described clear $\pi$* resonances (Fig.~\ref{fig:class3}a,b). Next are the spectra for a hypothetical model of the freestanding overlayer of molecules in their deformed adsorption geometry (Fig.~\ref{fig:class3}c,d). These spectra are vastly different from the gas-phase spectra, proving that a pure deformation of the molecule does have a significant effect on the spectral features. However, the spectra still contain distinct peaks with the energetic shifts for some transitions leading to even more separately distinguishable peaks. In the case of naphthalene, there is now a clear gap between the first two transitions. Therefore it is clear that the deformation is not the leading cause of the drastic change seen in the spectra caused by the adsorption on Pt (Fig.~\ref{fig:class3}g,h). 

To form a second hypothetical model, the undeformed, planar, gas-phase molecules were placed on the Pt surface in the correct adsorption sites and at the average adsorption height found in the equilibrium adsorption geometry. The corresponding calculated spectra (Fig.~\ref{fig:class3}e,f) are mostly featureless and indeed similar to the spectra of the fully relaxed adsorption geometries (Fig.~\ref{fig:class3}g,h). Therefore it is clear that interaction with the surface and not the deformation of the molecule forms the main reason for the massive change in the spectral features.

If we focus on the reasons for the diminished dichroism, however, the situation is a bit more nuanced. The residual $\pi$* intensity at \ang{90} incidence is present in similar magnitude for both the deformed, isolated molecule, and the undeformed molecule on the Pt surface (see Fig. S8 of the SI).

In summary, by calculating NEXAFS spectra of two hypothetical structures, we could clearly show that the main reason for the severe broadening and the loss of any distinct peaks in the spectra of azulene and naphthalene adsorbed on Pt(111) is the electronic hybridization due to interaction with the metal surface and not the deformation. For the loss of dichroism, however, both electronic hybridization and structural deformation are equally responsible.

\section{Discussion}

Through our simulated XP and NEXAFS spectra, we are able to provide a level of insight into the mechanisms at work in the spectroscopy of molecules adsorbed on metal surfaces that experimental techniques alone can not accomplish. In the following, we will generalize on our findings as a guide to identify regimes of interaction strength using calculated spectroscopic data of molecule/metal interfaces. A summary of our main identification criteria is presented in Table~\ref{tab:interaction_guide}. 

The physisorption regime describes weak molecule-metal coupling. Here, both the XPS and NEXAFS spectra of the adsorbed molecules show little change when compared to spectra of the gas-phase molecule. Whilst the absolute energies are different due to the presence of the metal surface, no change is seen in the XPS peak shape (Fig. S3) and for the NEXAFS spectra (Fig.~\ref{fig:np_class1} and S4), only a slight broadening is observable with all spectral features still distinctly identifiable. This behavior is an indication for the vanishing charge transfer as well as negligible hybridization of electronic states across the interface. Furthermore, the polarization dependence of the NEXAFS spectra is not affected and the expected dichroism of an ordered molecular monolayer on the surface remains intact. In such a case, the dichroism can relatively reliably be used to extract information about the molecular orientation.

The weak chemisorption regime shows significant interaction between the molecule and the metal. In the XPS calculations, we observe that molecule-metal interactions lead to changes in relative binding energy shifts between atoms. This effect may be expressed as a change of peak shape, as observed for the loss of the shoulder in the XP spectrum of azulene/Cu(111) (Fig.~\ref{fig:class2}a). In this specific case, the change in binding energies is caused by metal-to-molecule charge transfer. In the NEXAFS spectra, the most significant change compared to the gas-phase is the change in intensity of the leading peak(s). As the first signals typically correspond to the excitation into the lowest unoccupied orbitals, these peaks are important bellwethers for the interaction strength. In Az/Cu(111), the Fermi level pinning of the LUMO with the surface and the resulting charge transfer into the LUMO directly cause a reduction of the leading peak. In this bonding regime, dichroism in NEXAFS spectra is still visible, but dipolar selection rules for the transitions are softened. Due to this entanglement of electronic and structural factors, the analysis of the dichroism to determine the molecular orientation will likely be prone to errors.

In the strong chemisorption regime, the interaction between molecule and metal is a dominant influence in the spectroscopic data. In the XP spectra, the observed effects are similar to the weak chemisorption case. However, the changes in the NEXAFS spectra can provide a much clearer indication. Here we see a strong loss of intensity of the leading peak and extreme broadening across the whole spectrum resulting in a loss of distinct features. Despite the stark differences in the electronic structure of the gas-phase molecules, the spectra of our two model molecules are virtually indistinguishable when adsorbed on Pt(111). While the loss of intensity in the leading peak is similar to the case of weak chemisorption, it is not a sign of the same simple charge transfer mechanism. Instead, it is the expression of a complex charge donation and back-donation mechanism between the molecule and metal, accompanied by broadening of all involved MOs and their hybridization with the metal surface. The polarization dependence of the spectra is weak as most transitions include contributions from metal states. The complete breakdown of the selection rules as well as the deformation of the molecule make the analysis of the dichroism to extract structural information unfeasible.

\begin{table*}[ht!]
    \centering
    \caption{\label{tab:interaction_guide}Summary distinctive features depending on the interaction strength at the molecule/metal interface.}
    \begin{tabular}{lp{0.7\linewidth}}
    \toprule
   Bonding Regime               & Key features  \\
    \midrule
    \begin{tabular}[c]{@{}l@{}}Type I:\\Physisorption\end{tabular} & \begin{tabular}[c]{@{}p{\linewidth}@{}}\textbf{XPS features:} No significant changes to binding energies \\ \textbf{NEXAFS features:}  Spectrum similar to gas-phase with slight broadening\\\textbf{Dichroism:} Clear dichroism for $\pi^*$ resonances \\\textbf{Molecular Orbitals:}A clear HOMO-LUMO gap
    \end{tabular}    \\
    \midrule
    \begin{tabular}[c]{@{}l@{}}Type II:\\Weak Chemisorption\end{tabular} & \begin{tabular}[c]{@{}p{\linewidth}@{}}\textbf{XPS features:} Significant changes to relative binding energies of species in the molecule possible \\ \textbf{NEXAFS features:} Leading peak intensity reduced and significant broadening\\\textbf{Dichroism:} Weakened dichroism\\\textbf{Molecular Orbitals:} Loss of sharp HOMO and LUMO orbitals with LUMO partially below Fermi level
    \end{tabular}  \\
    \midrule
    \begin{tabular}[c]{@{}l@{}}Type III:\\Strong Chemisorption\end{tabular} & \begin{tabular}[c]{@{}p{\linewidth}@{}}\textbf{XPS features:} Similar to weak chemisorption\\ \textbf{NEXAFS features:} Loss of distinguishable spectral features. Loss of spectral differences between different molecules.\\\textbf{Dichroism:} Significantly reduced  dichroism\\\textbf{Molecular Orbitals:} Orbitals around the Fermi level completely smeared out, LUMO partially below Fermi level and HOMO paritally above
    \end{tabular}   \\
    \bottomrule
    \end{tabular}
\end{table*}


\section{Conclusion}

The aromatic topological isomers naphthalene and azulene adsorbed on the  (111) surfaces of Ag, Cu, and Pt present a group of model systems covering a wide reactivity range. Using these model systems, we conducted a systematic study of how the interaction strength at the molecule/metal interface influences core-level spectroscopy. The basis for this study is formed by XP and NEXAFS spectra calculated using state-of-the-art DFT methods. Good agreement was found between the calculation results and already published experimental measurements, allowing for a confident analysis of the computational data.

Our calculations allow us to probe the core-level spectra in ways not possible through experimental analysis alone. By decomposing both the XP and NEXAFS spectra into their initial or final state contributions, we can identify how different core and valence states contribute to the final spectra. Projection of the individual orbitals in NEXAFS spectra allows us to accurately assign spectral features and to assess the presence of molecule-surface charge transfer and its effect on the spectra. A great advantage of computational modeling is the ability to create experimentally unrealized structures to elicit information on how specific aspects such as molecular deformation and molecule-surface interaction contribute to changes in spectra.

Overeall, we identified three regimes of interaction: Physisorption [systems: Nt/Ag(111), Nt/Cu(111), Az/Ag(111)], weak chemisorption [Az/Cu(111)], and  strong chemisorption [Nt/Pt(111), Az/Pt(111)]. By careful analysis, we were able to pinpoint specific markers to justify classification into the different interaction regimes. A summary of the markers present in both the XPS and NEXAFS spectra for each regime has been provided in Table~\ref{tab:interaction_guide} of the discussion. By understanding the spectral change in different molecule-surface interaction scenarios, important insights on the chemical bonding and charge transfer at the surface can be gained. We expect that our findings can be generalized to many other systems, in particular to conjugated organic molecules or nanographene on metal surfaces.

\begin{acknowledgement}
\noindent
This work is, in parts, based on chapter 5 of Samuel J. Hall's dissertation \cite{hall_computational_2022}. S.J.H and R.J.M acknowledge funding for a PhD studentship through the EPSRC Centre for Doctoral Training in Molecular Analytical Science (EP/L015307/1) and computing resources via the EPSRC-funded HPC Midlands+ computing centre (EP/P020232/1) and the EPSRC-funded Materials Chemistry Consortium for the ARCHER2 UK National Supercomputing Service (EP/R029431/1). B.P.K. is supported by  Deutsche Forschungsgemeinschaft (DFG, German Research Foundation) via grant KL 3430/1-1. R.J.M acknowledges support via a UKRI Future Leaders Fellowship (MR/S016023/1). The authors thank Michael Gottfried and Ralf Tonner-Zech for fruitful discussions.

\end{acknowledgement}

\section{Associated Content}
\subsection{Supporting Information}
Experimental comparison of simulated core-level spectra, charge transfer  analysis, x-ray photoemission spectra, molecular orbital projected NEXAFS spectra, molecular orbital projected density of states, NEXAFS spectra dichroism.

\section*{Conflicts of Interest}
\noindent
The authors declare no conflict of interest.

\section*{Author Information}
\subsection{Corresponding Author}
\noindent
Reinhard J. Maurer, r.maurer@warwick.ac.uk
\subsection{ORCIDs}
\noindent
Samuel J. Hall: 0000-0003-3765-828X \\
Benedikt P. Klein: 0000-0002-6205-8879\\
Reinhard J. Maurer: 0000-0002-3004-785X\\

\section*{Data Availability Statement}
All input and output data files of the calculations performed and presented in this publication are freely available online in the NOMAD electronic structure data repository at the following under: https://dx.doi.org/10.17172/NOMAD/2021.06.14-1.

\bibliography{bibliography}

\end{document}



\clearpage

\section{Supplementary Information}

\begin{figure}[b!]
    \includegraphics{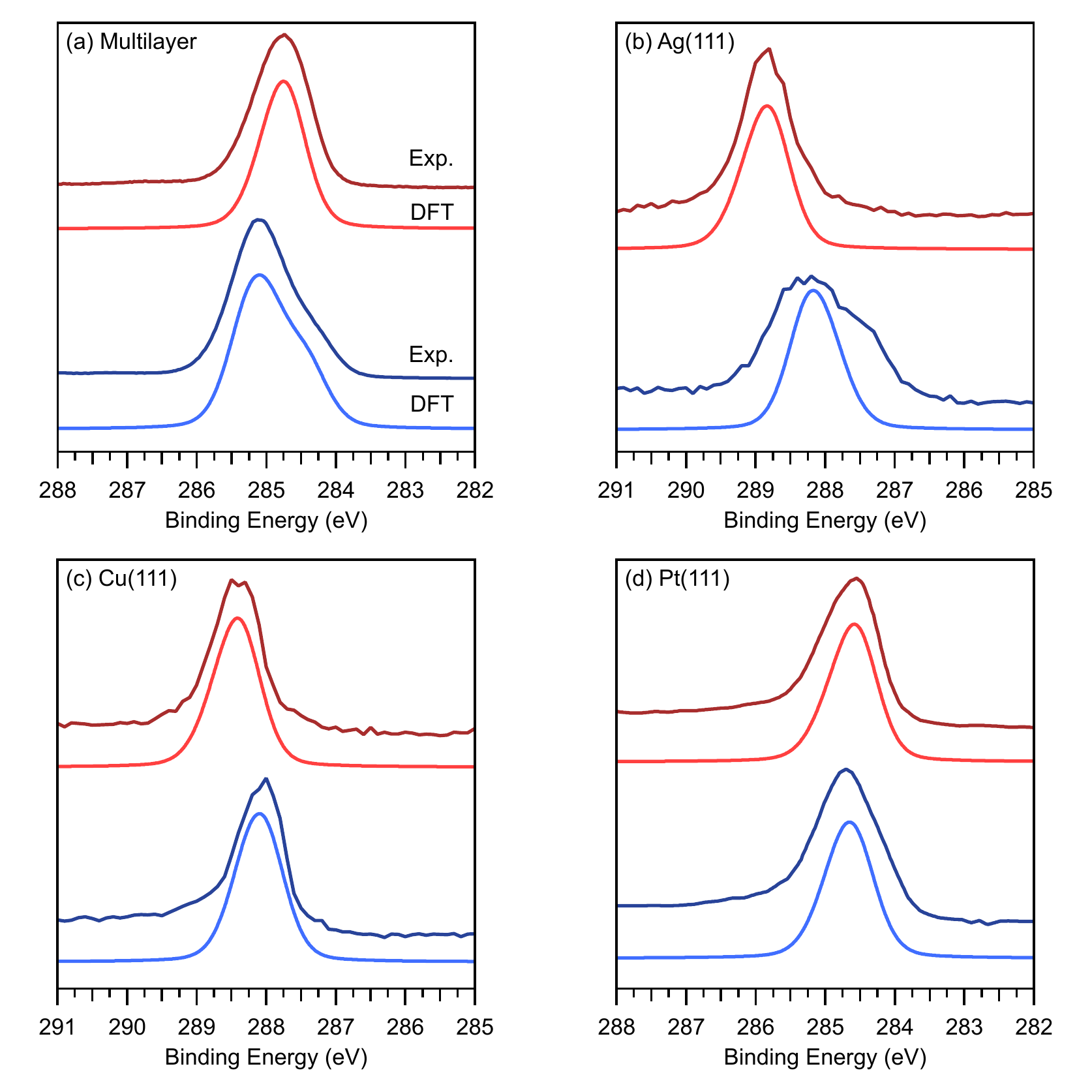}
    \caption{\label{fig:exp_xps_comp}Comparison of computational results to experimentally recorded C 1s XPS spectra of naphthalene (in red) and azulene (in blue) in a multilayer sample (a), adsorbed on an Ag(111) surface (b), on Cu(111) in (c), and on Pt(111) in (d). Experimental data previously published in Refs.~\citenum{klein_molecular_2019,klein_molecule-metal_2019,klein_enhanced_2020}. Experimental spectra shown on top in darker shade whilst below in lighter shade are the DFT simulated results. A shift was applied to the computational results to match the experimental energy scale.}
\end{figure}

\begin{figure}[t!]
    \centering
    \includegraphics[width=0.85\linewidth]{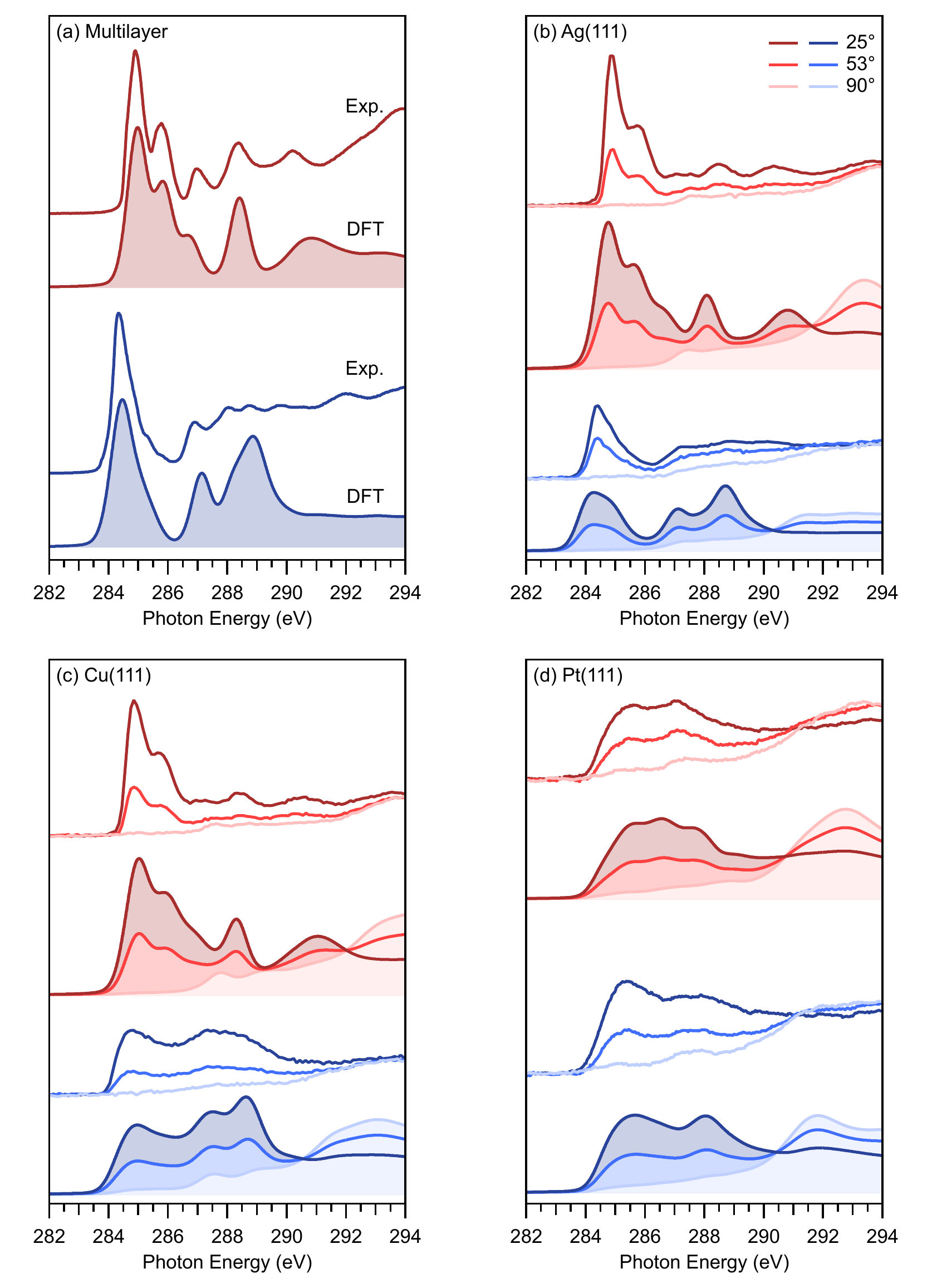}
    \caption{\label{fig:exp_nexafs_comp}Comparison of computational results to experimentally recorded C K-edge NEXAFS spectra of naphthalene (red) and azulene (blue) previously published in Refs.~\citenum{klein_molecular_2019,klein_molecule-metal_2019,klein_enhanced_2020}. (a) shows the spectra of a multilayer sample recorded at a \ang{25} incidence angle. (b) are spectra for the molecule adsorbed on an Ag(111) surface, (b) adsorbed on Cu(111), and (d) adsorbed on Pt(111). Spectra for three different incidence angles of \ang{25}, \ang{53}, and \ang{90} are shown from darker to lighter line shades. Experimental spectra are shown at the top with simulated DFT spectra below as shaded spectra. A shift was applied to the computational results to match the experimental energy scale.}
\end{figure}

\clearpage

\begin{table}[t!]
    \centering
    \caption{\label{tab:charges} Calculated net charge transfer of all metal adsorbed systems investigated using various charge analysis methods and electronic structure codes . All values are given in e, a negative value means electrons are transferred from the surface to the molecule. Bader and DOS charges for all systems, as well as Hirshfeld values for Pt systems, have previously been reported in Refs.~\citenum{klein_molecule-metal_2019,klein_enhanced_2020}.}
    \begin{tabular}{c|cccccc}
    \toprule
    Method             & \mbox{Nt/Ag} & Az/Ag & Nt/Cu & Az/Cu & Nt/Pt & Az/Pt \\
    \midrule
    DOS (CASTEP)        & -0.05 & -0.21 & -0.13 & -1.39 & -1.70 & -1.60 \\
    Hirshfeld (CASTEP)  & -0.07 & -0.11 & -0.05 & -0.35 & -0.32 & -0.31 \\
    Hirshfeld (VASP)    & -0.04 & -0.07 & -0.01 & -0.25 & -0.20 & -0.21 \\
    It-Hirshfeld (VASP) & -0.14 & -0.19 & -0.12 & -0.60 & -0.86 & -0.84 \\
    Bader (ADF-BAND)  &  +0.06 &  +0.01 &  +0.06 & -0.49 &  +0.03 &  +0.02 \\
    \bottomrule
    \end{tabular}
\end{table}

\begin{figure}[b!]
    \centering
    \includegraphics{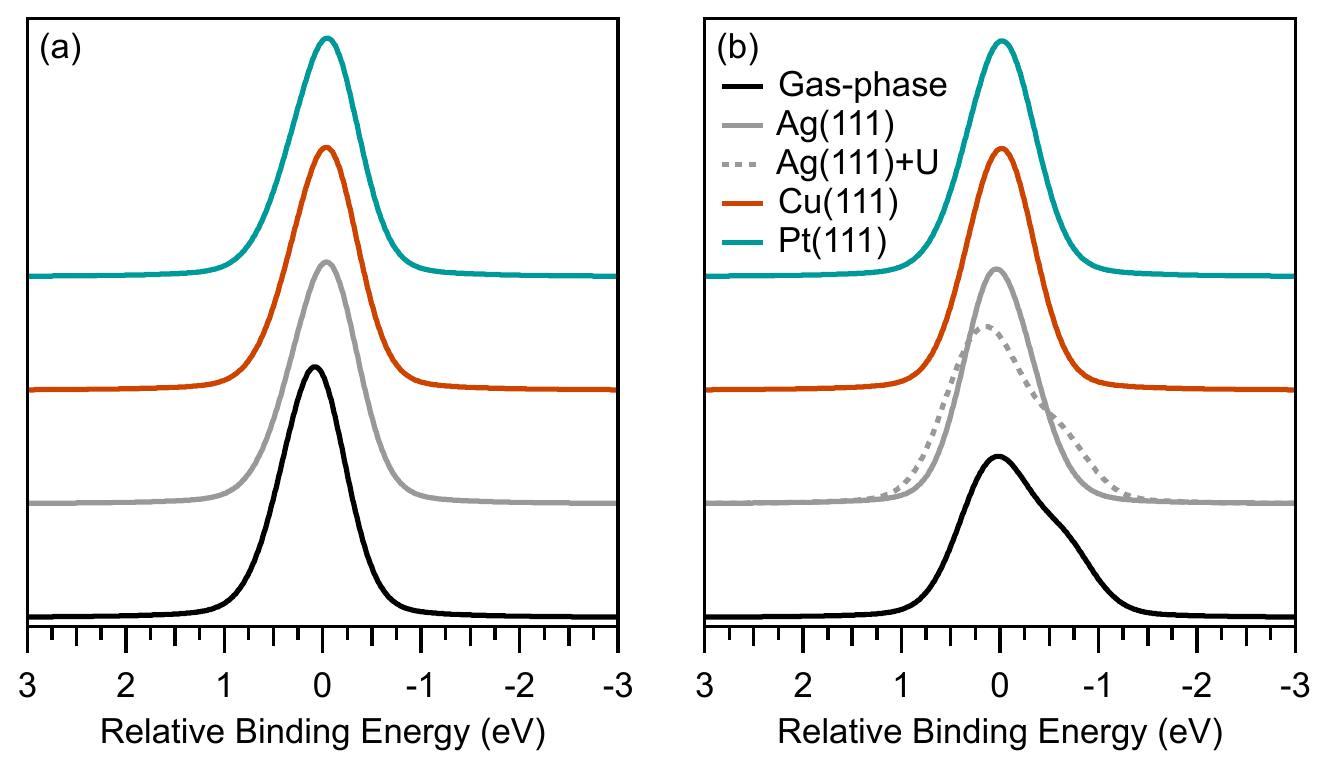}
    \caption{\label{fig:xps_compare}Comparison of XPS spectra for naphthalene (a) and azulene (b) in the gas-phase (black) previously published in Ref.~\citenum{klein_molecular_2019}, and adsorbed on three metal surfaces, Ag(111) (gray), Cu(111) (orange) and Pt(111) (green). Spectra of the metal adsorbed systems have been aligned to the center of mass.}
\end{figure}

\clearpage

\begin{figure}[t!]
    \centering
    \includegraphics{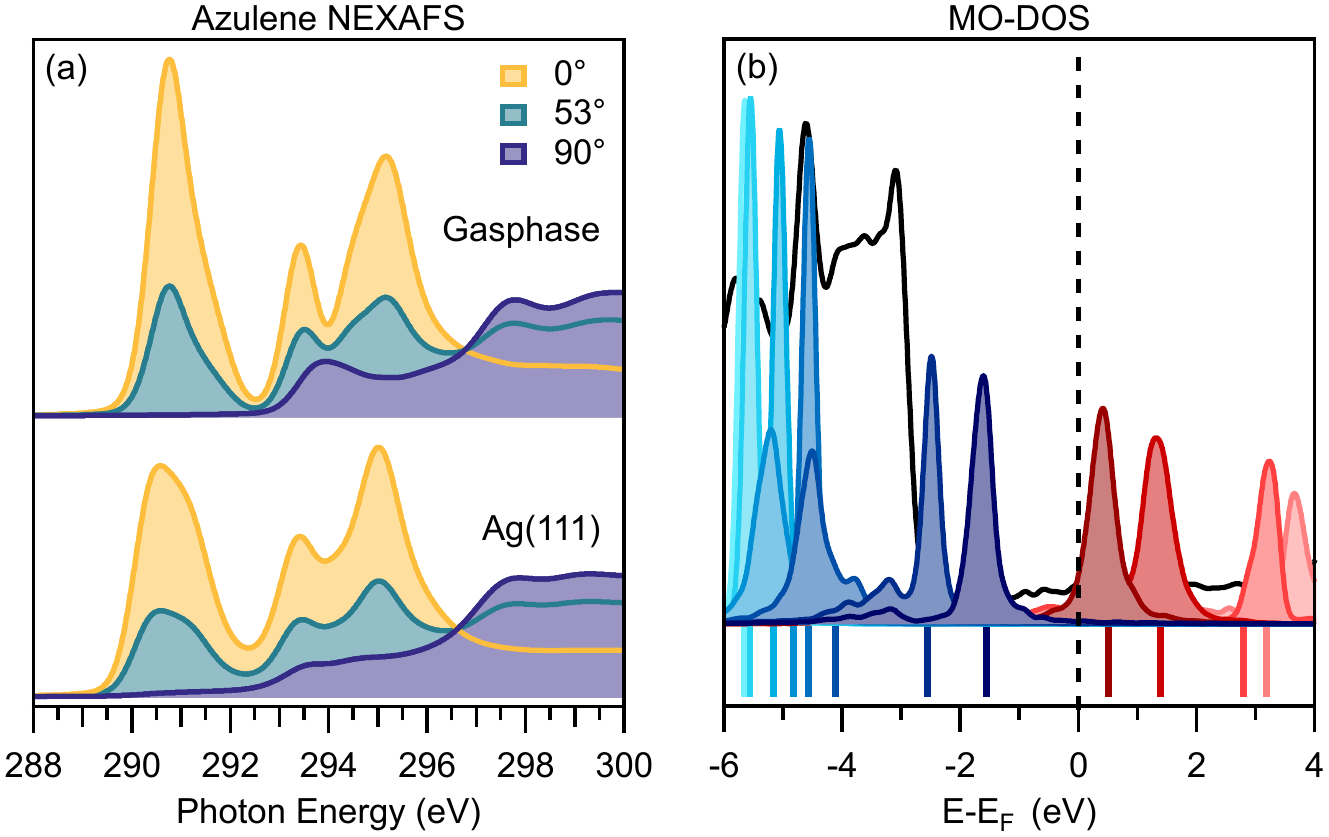}
    \caption{\label{fig:Az_class1}(a) DFT calculated NEXAFS spectra of azulene in two different systems. Gas-phase (top) and the physisorbed Az/Ag(111) (bottom). Three different incidence angles are shown of \ang{0} (yellow), \ang{53} (green) and \ang{90} (blue). (b) shows DFT density of states (DOS) of the metal adsorbed system. Total DOS is shown in black and Fermi level shown with dashed line, contributions from orbitals scaled for ease of viewing gas-phase. Labeled in blue orbitals represent the HOMOs and red the LUMOs with the lighter shade moving lower or higher in energy respectively. Colored lines at the bottom of the graph represent gas-phase orbitals shifted by \SI{3.3}{\electronvolt} to aligned with the metal adsorbed system. All data was previously published in Ref.~\citenum{klein_molecule-metal_2019}, except the \ang{0} NEXAFS spectra.}
\end{figure}

\begin{table}[b!]
    \centering
    \caption{\label{tab:azulene}Absolute energies and intensities of first 3 peaks seen in NEXAFS spectra of azulene gas-phase and on Ag(111) and Cu(111) surfaces for spectra simulated with a \ang{0} incidence angle. All intensities have been normalized to the leading Az/Gas peak.}
    \resizebox{\textwidth}{!}{
    \begin{tabular}{c|cc|cc|cc}
    \toprule
    \multirow{2}{*}{Peak} & \multicolumn{2}{c}{Az/Gas}     & \multicolumn{2}{c}{Az/Ag(111)} & \multicolumn{2}{c}{Az/Cu(111)}  \\
                      & Photon Energy (eV) & Intensity & Photon Energy (eV) & Intensity & Photon Energy (eV) & Intensity  \\
    \midrule
    1 & 290.77 & 1.00 & 290.58 & 0.65 & 290.97 & 0.43 \\
    2 & 293.43 & 0.48 & 293.42 & 0.45 & 293.51 & 0.50 \\
    3 & 295.17 & 0.73 & 295.02 & 0.70 & 294.62 & 0.59 \\
    \bottomrule
    \end{tabular}}
\end{table}

\begin{table}[b!]
    \centering
    \caption{\label{tab:azulene_rel)heights}Relative intensities of the first three peaks in the NEXAFS spectrum of azulene in the gas-phase, Az/Ag(111), and Az/Cu(111), normalised to the intensity of the third peak in each spectrum.}
    \begin{tabular}{c|ccc}
    \toprule
    Peak & Az/Gas & Az/Ag(111) & Az/Cu(111) \\
    \midrule
    1 & 1.37 & 0.93 & 0.73 \\
    2 & 0.66 & 0.64 & 0.84 \\
    3 & 1.00 & 1.00 & 1.00 \\
    \bottomrule
    \end{tabular}
\end{table}

\begin{table}[b!]
    \centering
    \caption{\label{tab:azulene_rel)heights2}Relative intensities of the first two peaks in the NEXAFS spectrum of Az/Ag(111) and Az/Cu(111), first normalised to the third peak in each spectrum then to the intensity of the corresponding peak intensity of gas-phase azulene (Table~\ref{tab:azulene_rel)heights}).}
    \begin{tabular}{c|cc}
    \toprule
    Peak & Az/Ag(111) & Az/Cu(111) \\
    \midrule
    1  & 0.68 & 0.53 \\
    2  & 0.97 & 1.27 \\
    3  & 1.00 & 1.00 \\
    \bottomrule
    \end{tabular}
\end{table}

\begin{figure}[t!]
    \includegraphics{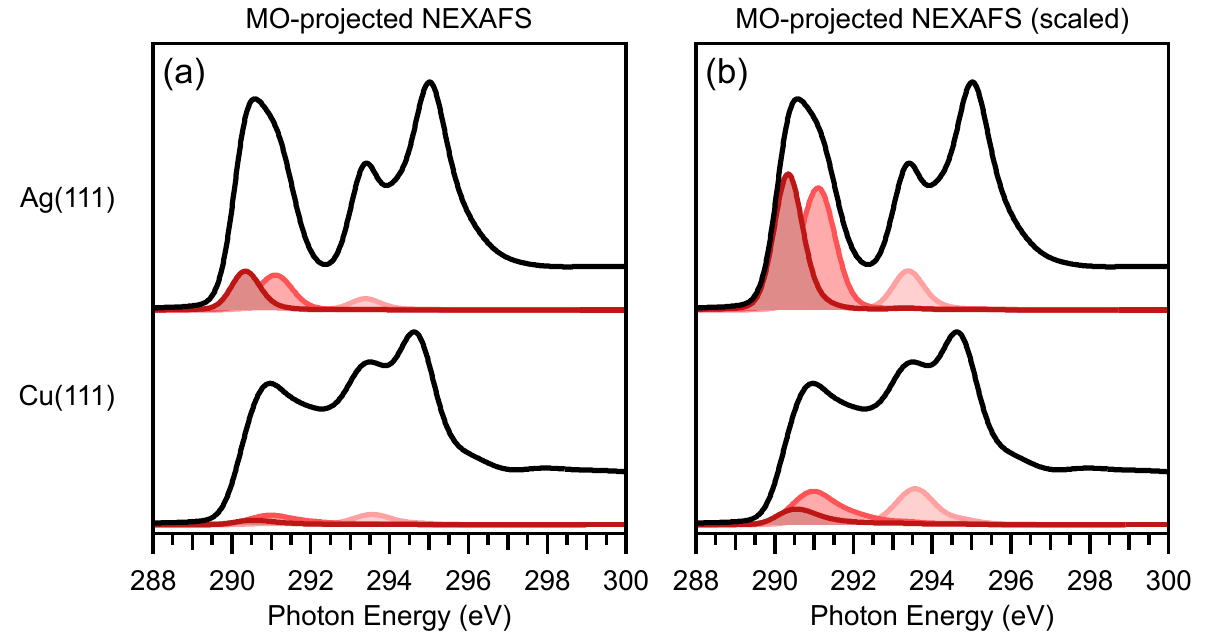}
    \caption{\label{fig:class2_MO_contri}MO-projected NEXAFS of Az/Ag(111) (top) and Az/Cu(111) (bottom) using the simulated \ang{0} spectra. (a) presents the MO-decomposition as calculated and (b) scales contributions for both Ag(111) and Cu(111) by the same amount for ease of viewing.}
\end{figure}

\begin{figure}[t!]
    \includegraphics{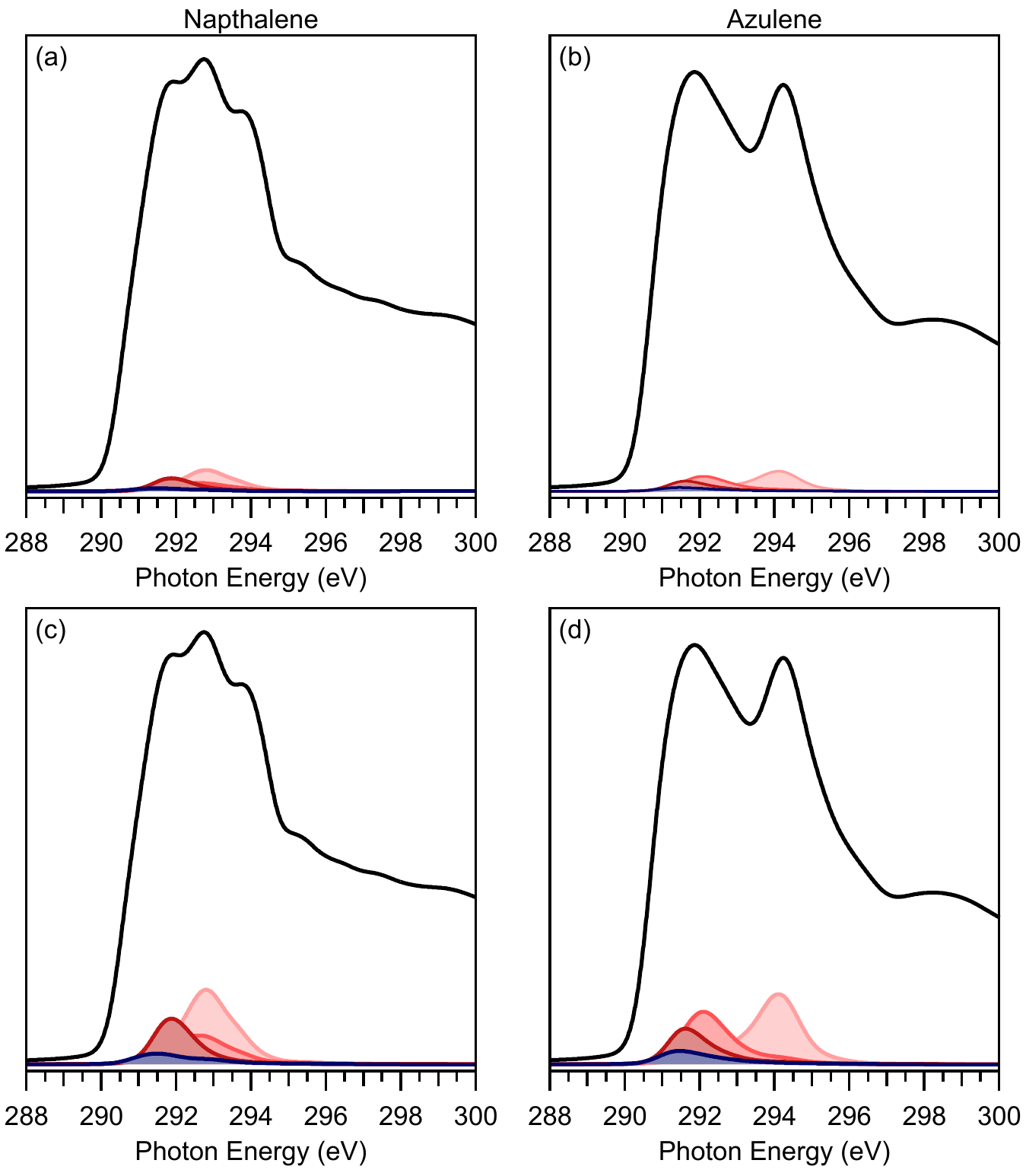}
    \caption{\label{fig:class3_MO}MO decomposition of \ang{0} NEXAFS of the type III systems of naphthalene, (a), and azulene, (b), adsorbed on Pt(111). (c) and (d) show the same decomposition as above but with the MO orbitals scaled for ease of viewing. Previously published in Ref.~\citenum{klein_enhanced_2020}, recalculated with updated methodology.}
\end{figure}

\begin{figure}[t!]
    \centering
    \includegraphics{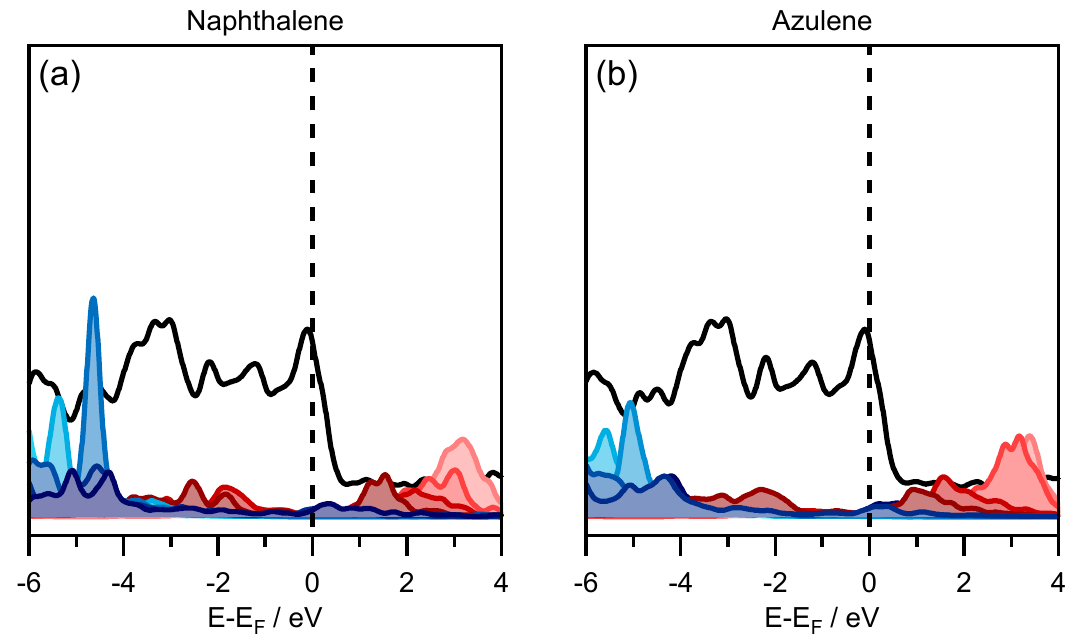}
    \caption{\label{fig:class3_DOS}DFT density of states DOS  with scaled MO projections for naphthalene (a) and azulene (b) adsorbed on a Pt(111) surface. The total DOS is shown in black with Fermi level shown as dashed vertical line. Contributions in blue represent projected occupied states, while contributions in red represent unoccupied states. Data previously published in Ref.~\citenum{klein_enhanced_2020}.}
\end{figure}

\begin{figure}[b!]
    \includegraphics[width=\linewidth]{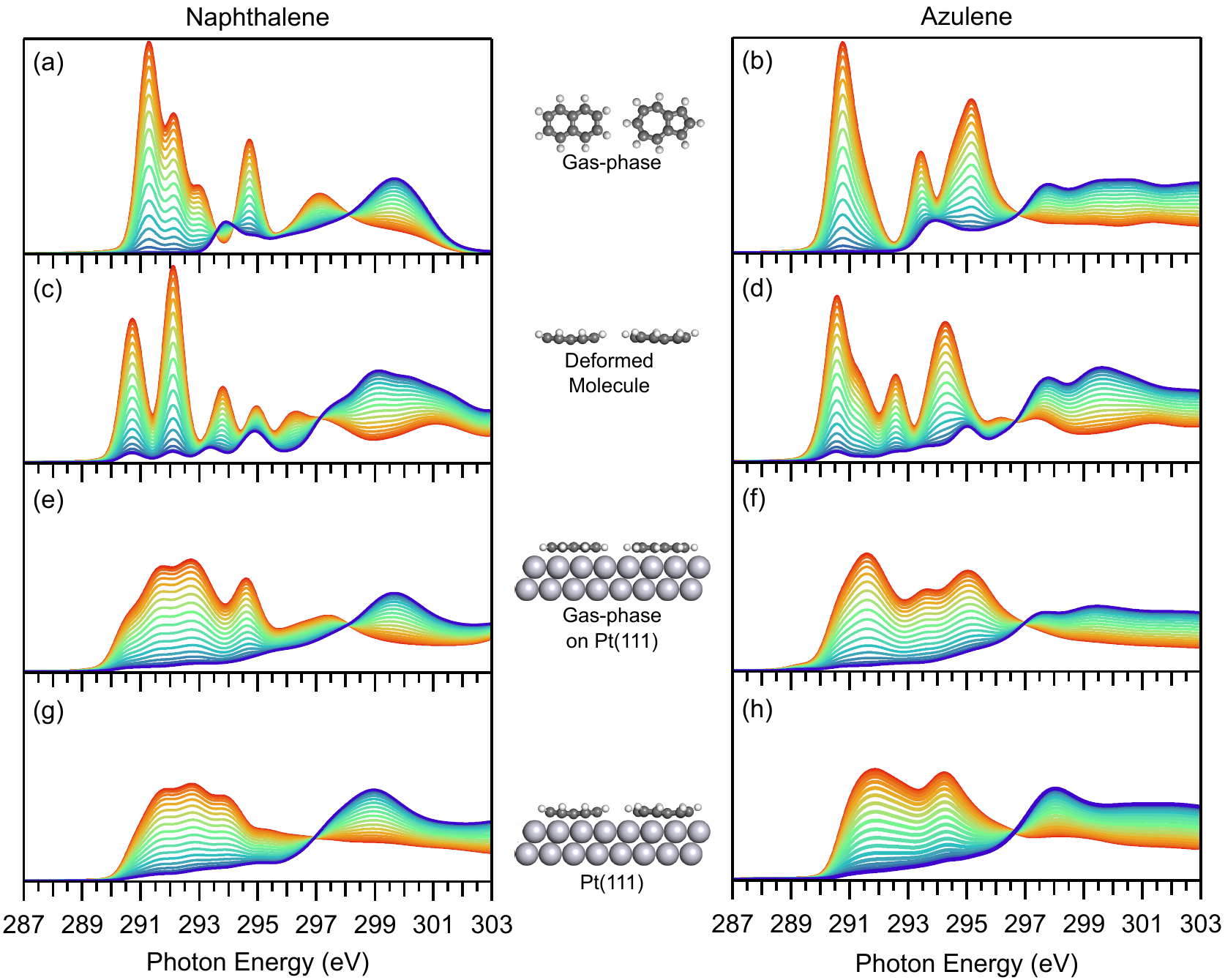}
    \caption{\label{fig:class3_angle}Simulated NEXAFS spectra of naphthalene (left) and azulene (right). Dichroism is depicted by showing spectra with varying incidence angles from normal incidence \ang{90} (blue) to grazing incidence \ang{0} (red). (a,b) show the spectra for the gas-phase molecule, (c,d) for the molecule in the gas phase, but with the adsorption induced deformation, (e,f) for the planar gas-phase molecule adsorbed on Pt with the proper adsorption site and height, and (g,h) for the fully relaxed structure of the molecule adsorbed on Pt. Next to all spectra is a model visualizing the corresponding structure.}
\end{figure}

\clearpage
\bibliography{bibliography}